\documentclass[a4paper,11pt]{article}
\pdfoutput=1 % if your are submitting a pdflatex (i.e. if you have
\usepackage{jheppub,} % for details on the use of the package, please
\usepackage{tikz}
\usetikzlibrary{decorations.pathmorphing}
\usetikzlibrary{shadows.blur}

\usepackage[T1]{fontenc} % if needed
\usepackage{braket}
\def\Tr{\mathrm{Tr}}

\def\TFD{\mathrm{TFD}}
\def\be{\begin{equation}}
\def\ee{\end{equation}}
\def\ra{\rangle}
\def\la{\langle}
\def\hat{\widehat}
\newcommand{\BUket}[6]{%
 \begin{scope}[shift={(#1,#2)}, yscale={\ifnum#3=1 -#4\else #4\fi}, xscale=#4]
 \ifnum#6=1
 \draw[dotted, thick] (0,0) ellipse (.5 and .25);
 \draw[dotted, thick] (3,0) ellipse (.5 and .25);
 \draw[thick] (0.5,0) arc[start angle=0, end angle=180, radius=0.5];
 \draw[thick] (3.5,0) arc[start angle=0, end angle=180, radius=0.5];
 \else
 \draw[thick] (0,0) ellipse (.5 and .25);
 \draw[thick] (3,0) ellipse (.5 and .25);
 \fi

 \ifcase#5
 \draw[thick,dashed] (1.5,0) ellipse (.3 and .15); % 0 = dashed
 \or
 \draw[double,thick] (1.5,0) ellipse (.3 and .15); % 1 = thick double
 \or
 \draw[very thick,green] (1.5,0) ellipse (.3 and .15); % 2 = thick green
 \or
 \draw[thick] (1.5,0) ellipse (.3 and .15); % 3 = normal solid
 \fi

 \draw[thick] (3.5,0) arc[start angle=0, end angle=-180, radius=2];
 \draw[thick] (2.5,0) arc[start angle=0, end angle=-180, radius=1];
 \draw[thick, dashed] (1.2, 0) to[out = -90, in =80] (1,-1.5);
 \draw[thick, dashed] (1.8, 0) to[out = -90, in =100] (2,-1.5);
 \end{scope}%
}

\title{Emergent Mixed States for Baby Universes and Black Holes}

\author[]{Jonah Kudler-Flam and}
\author[]{Edward Witten}

\affiliation[]{School of Natural Sciences, Institute for Advanced Study, Princeton, NJ, USA}

\emailAdd{jkudlerflam@ias.edu}
\emailAdd{witten@ias.edu}

\abstract{We examine the behavior of sequences of states in the large $N$ limit of AdS/CFT duality in cases in which the bulk duals involve baby universes or black holes. Such sequences generally fail to converge as pure states. 
Under suitable conditions, such as diverging coarse-grained entropy, they can
converge to mixed states for the large $N$ algebra, as in the case of black holes. For Euclidean preparations that produce baby universes, the sequences do not converge, due to wormhole contributions, and so these states cannot admit large $N$ limits. Nevertheless, appropriate averaging over $N$ can lead to convergence to a mixed state. The associated algebras have nontrivial commutants, which can possibly be interpreted as operators in the baby universe.}

\begin{document} 
\maketitle
\flushbottom
\def\A{{\mathcal A}}
\def\var{{\mathrm {var}}}
\def\OO{{\Bbb O}}
\section{Introduction}
\label{sec:intro}

The AdS/CFT correspondence proposes exact dualities between non-gravitational conformal field theories and string theories \cite{Maldacena:1997re,Gubser:1998bc,Witten:1998qj}. However, the vast majority of its study has come from taking the large $N$ (and large 't Hooft coupling) limit of the conformal field theory, where the number of degrees of freedom scales as some power of $N$. This corresponds to the supergravity limit where Newton's constant, $G$, is small and quantum field theory in curved spacetime is a good approximation. The large $N$ limit is subtle 
 because not all operators survive the limit and the operators
that do survive do not always suffice to fully describe the state. This leads to a sort of information loss. For example, the algebra of observables that survive the limit, when considering one side of the thermofield double state at temperatures above the Hawking-Page transition, is a type III$_1$ von Neumann algebra \cite{Leutheusser:2021frk,Leutheusser:2021qhd,Furuya:2023fei}, an algebra that admits no pure states, even though for every finite value of $N$, the algebra is of type I and admits pure states.

\begin{figure}
 \centering
 \begin{tikzpicture}
 \draw[ blur shadow={shadow blur steps=3,
 draw=none, shadow xshift=0pt,
 shadow yshift=0pt,}] (0,0) -- (2,2) -- (0,4) -- (2,2) ;
 \draw[thick] (0,0) to[out = 0, in = -90] (3,2)to[out = 90,in = 0] (0,4)--cycle;
 \filldraw[gray, opacity = .25] (0,0) to[out = 0, in = -90] (3,2)to[out = 90,in = 0] (0,4);
 \end{tikzpicture}
 \hspace{3cm}
 \begin{tikzpicture}
 \filldraw[gray, opacity = .25] (0,0) -- (0,4) -- (2,2) --cycle;
 \draw[
 thick
 ] (0,0) -- (2,2) -- (0,4) --cycle ;
 \draw[decorate, decoration={zigzag}, thick] (0,0) -- (4,0);
 \draw[decorate, decoration={zigzag}, thick] (0,4) -- (4,4);
 \end{tikzpicture}
 \caption{Left: At finite $N$, the high-energy state is pure and then dual to the entire bulk, whose geometry has quantum fluctuations and so there is no sharp horizon. Right: The limit of a sequence of high-energy pure states is dual to the mixed state in the gray bulk region.}
 \label{fig:onesided}
\end{figure}
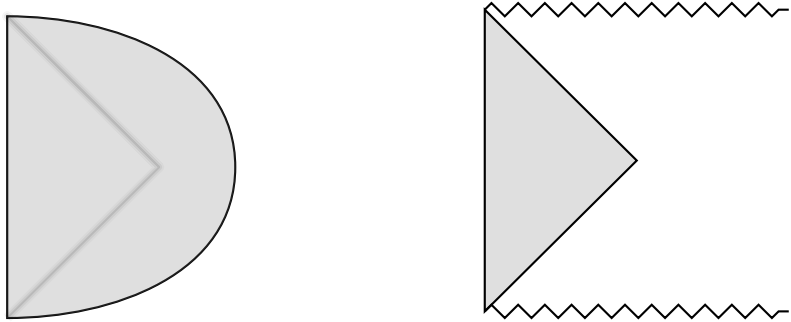

In this article, we investigate how sequences of states behave in the large $N$ limit. The first thing to note is that there is no simple relation between the Hilbert space of states at gauge group rank $N$ and $N+1$ and so the Hilbert space itself does not have a large $N$ limit. 
However, there are sectors of the Hilbert space that do have large $N$ limits, the most obvious example being the low-energy sector spanned by states obtained by acting with a finite number of single-trace operators on the vacuum vector. On the gravity side, this sector consists of excitations of Anti de Sitter
(AdS) space with $O(N^0)$ energy. In contrast, we may consider a sequence of high-energy states of the conformal field theory, with energy scaling with $N$ in such a way that the energy is a fixed multiple of $1/G$ (for ${\mathcal N}=4$
super Yang-Mills theory, this means that the energy scales as $N^2$). 
As mentioned, there is generally no canonical way to relate states at different values of $N$, but we can imagine picking,
for example, any sequence of energy eigenstates whose energy scales with $N$ in the stated fashion.
Such a sequence of states will not converge to a vector in some large $N$ Hilbert space. Nevertheless, if one considers only an algebra of (properly normalized) single-trace operators of appropriate energy, under the assumption of the eigenstate thermalization hypothesis \cite{Srednicki:1995pt}, all expectation values will converge in the large $N$ limit 
to expectations in a thermal ensemble and so the sequence of algebraic states\footnote{An algebraic state, $\omega$, is a positive linear functional on the algebra with $\omega(1) = 1$.} converges. The key difference from the vacuum sector is that this sequence converges to a mixed state, namely the thermal state. 
Thus, although single-trace operators are believed to generate the full operator algebra for any fixed value of $N$,
in the large $N$ limit there can be information loss if the state is probed by the algebra of single-trace operators.
This information loss in the boundary theory manifests in the gravity theory by the existence of a black hole
horizon; the region beyond the horizon is not accessible to the algebra of single-trace operators in the large
$N$ limit (see fig.~\ref{fig:onesided}) \cite{Leutheusser:2021frk,Leutheusser:2021qhd, Engelhardt:2023xer, Gesteau:2024rpt, Liu:2025krl}. All information about the interior is lost in the large $N$
 limit, assuming
 that one only has access to the algebra of single-trace operators.
 
In the large $N$ limit, there is also a third option that probes a new regime.
Here we are motivated by candidate
gravitational saddles described by\footnote{
This work built on prior literature in which operators creating shells of matter were introduced to support wormhole geometries \cite{Engelhardt:2018kcs, Sasieta:2022ksu, Chandra:2022fwi,Balasubramanian:2022lnw,Balasubramanian:2022gmo}. These shell operators were previously used as models of black hole collapse \cite{Anous:2016kss}.}  Antonini, Sasieta, and Swingle (AS$^2$)  \cite{Antonini:2023hdh} 
and a puzzle about these saddles that was 
 emphasized by Antonini and Rath (AR) \cite{Antonini:2024mci}.
 The saddles in question involve Euclidean preparation, in the AdS/CFT context,
 of what appears to be a joint state of two
 asymptotically AdS regions entangled with a closed baby universe. AS$^2$ argued that these bulk
 states are dual to 
 partially entangled thermal states \cite{Goel:2018ubv} of two copies of the dual CFT.
 The puzzle is that from a boundary point of view, the AS$^2$ construction apparently leads to an entangled
 pure state of two copies of the CFT with $O(N^0)$ energy and entropy, but from a bulk point of view, it appears
 that the two AdS regions are entangled with a baby universe and thus must be in a mixed state.
 One potential resolution of the puzzle considered in \cite{Antonini:2024mci} and
 further explored in \cite{Engelhardt:2025vsp,Higginbotham:2025dvf,Engelhardt:2025azi} relied on the fact that the gravitational path integral seems to imply that the Hilbert space dimension of baby universes is one-dimensional \cite{Almheiri:2019hni,Penington:2019kki,Marolf:2020xie}, at least for some purposes.
 This would potentially resolve the puzzle, because a one-dimensional Hilbert space cannot entangle with any other
 system.

In this paper, we resolve the puzzle by showing that the sequence of pure states does not converge in the large $N$ limit,\footnote{This has been argued independently in \cite{Liu:2025cml}.} due to fluctuations in expectation values of simple operators. 
Our main observation is based on standard reasoning about wormholes, as developed in
 \cite{Coleman:1988cy,Maldacena:2004rf}, for example.   We argue 
 that the very hypothesis that
a Euclidean state preparation produces, in the bulk description,
a state containing a baby universe implies that the associated 
 CFT state does not have a large $N$ limit as a pure state. This claim holds for the specific AS$^2$ construction
 involving two copies of the CFT, but more generally for any Euclidean state preparation involving a single copy
 of the CFT or any number of copies.   This simple observation resolves the AR paradox, which involves reasoning about the presumed large $N$ pure state limit of two copies of the CFT.

Though the CFT state does not have a large $N$ limit as a pure state, one can ask if it can have a limit as a mixed
state.\footnote{In \cite{Gesteau:2025obm}, it was shown that if sequences of finite energy states in AdS/CFT have a limit, the limiting state must be pure. Relatedly, in \cite{dell1967limits}, it was shown that sequences of pure states for type I von Neumann algebras can only converge to pure states, though the same does not hold for C$^\star$ algebras.} This is possible at least in toy models, as we explain in section \ref{two}, but only if the entanglement between the baby universe and the asymptotically AdS
regions diverges for $N\to\infty$. (In the literature on the AS$^2$ construction, this entanglement is usually
assumed to be finite. See \cite{Antonini:2025ioh} for a phase diagram including a portion where the entropy diverges.) The emergence here of a CFT mixed state as the large $N$ limit of a sequence of pure states with
divergent entanglement
is analogous to what happens in the example of black hole microstates discussed previously.

If the entanglement of the baby universe is finite, as usually assumed,
the CFT state does not have a large $N$ limit even as a mixed state. 
However, one can average the CFT density matrix over $N$, and, depending on details of the AS$^2$ construction or
of a toy model,
this average may converge to a mixed state of the large $N$ algebra of single-trace operators.
If so, the GNS construction can be applied to this limiting mixed state and will give a limiting large $N$
Hilbert space with an action of the
the large $N$ algebra of single-trace operators. Because the state is mixed, the algebra will have a nontrivial 
commutant and it may be reasonable to interpret these operators as operators in the baby universe.

We first analyze simple toy models in section \ref{two}, and then we consider gravity and the 
AdS/CFT correspondence in section \ref{three}. We conclude in section \ref{four} with comments on averaging over $N$.

\section{Qubit Models of Baby Universes and Black Holes}\label{two}

To gain intuition for how states can behave in the large $N$ limit of AdS/CFT, it is useful to consider some simple qubit models. In this section, we introduce and analyze some $N$-qubit states that we claim in the large $N$ limit faithfully model (i) the AS$^2$ set-up, (ii) one-sided black holes, and (iii) two-side black holes connected by a long wormhole.

We consider a family of systems labeled by a positive integer $N$, with ``simple operators''
$a_{i,N}$ that are expected to converge for $N\to\infty$ to limiting simple operators $a_i$.
A sequence of states $\{\omega_N\}$ for the $N^{th}$ system converges to a limiting
state $\omega$ if, for all $i$, 
\begin{align}\label{limitstate}
 \lim_{N\rightarrow \infty}\omega_N(a_{i,N}) = \omega(a_i).
\end{align}
Let $\mathcal{A}$ be the algebra generated by the $a_i$.
With this limiting algebra and with
the limiting state $\omega$ at hand, the GNS construction provides a Hilbert space $\mathcal{H}$ on which $\A$ acts. Taking the double commutant of $\mathcal{A}$ with respect to $\mathcal{H}$ leads to a von Neumann algebra, which we will also call $\mathcal{A}$.

There are three potential outcomes:
\begin{enumerate}
 \item The limiting state $\omega$ does not exist.
 \item The limiting state is a pure state, i.e.~$\mathcal{H}$ is an irreducible representation of $\mathcal{A}$.
 \item The limiting state is mixed, i.e.~$\mathcal{A}$ has a nontrivial commutant on $\mathcal{H}$.
\end{enumerate}

If the limiting state does not exist, one can consider some sort of averaging over $N$ to define
an average limiting state.
One simple way to average in $N$ is to consider the limit of partial sums
\begin{align}
 \lim_{N\rightarrow \infty}\frac{1}{N}\sum_{n = 1}^N\omega_n(a_{i,N})= \omega(a_i).
 \label{eq:partialsum}
\end{align}
This can turn a non-convergent sequence into a convergent one.
(As discussed by Liu \cite{Liu:2025cml}, many variants of this simple averaging procedure can be envisaged.)
If the original sequence already converges, then the averaging does not change the limiting state. 
If the original sequence does not converge, the average does not converge to a pure state but may 
nonetheless converge to a mixed state. In that situation, the GNS construction can still be implemented to define a Hilbert space with an action of a von Neumann algebra $\A$ that is a completion of the
algebra generated by the original simple operators $a_i$.

Depending on the limiting state, $\mathcal{A}$ can take different ``types.'' 
If the limit is a pure state, $\mathcal{A}$ is a type I von Neumann algebra. 
If the limit is mixed, the algebra can be type I, II, or III. 

\subsection{A Qubit Model of Baby Universes and Long Wormholes\label{babymodel}}

We can model the AS$^2$
state by taking the ``heavy'' operator used in that construction 
to be a random $2^N \times 2^N$ unitary\footnote{Here unitarity of $U_N$ is actually not essential.
For example, we could construct a somewhat similar toy model taking $U_N$ to be simply a complex Gaussian random matrix.} matrix
$U_N$
\begin{align}
 \ket{\psi_N} = \mathcal{N}^{-1/2} e^{-\beta_L H_N/2}U_Ne^{-\beta_R H_N/2}, \quad \mathcal{N} \equiv \Tr(e^{-\beta_L H_N}U_Ne^{-\beta_R H_N}U_N^{\dagger}),
\end{align} where $\mathcal N$ is a normalization factor. An operator acting on $N$ qubits can be viewed as a pure state of a doubled system
with $2N$ qubits, and we will interpret the state $\ket{\psi_N}$ in this way. 

First, let us consider the coefficients of this state when expanded in a basis for left and right systems
\begin{align}\label{normalizedstate}
 M_{ij} \equiv \bra{i}_L\bra{j}_R\ket{\psi_N} = \frac{1}{\mathcal{N}} \bra{i}e^{-\beta_L H_N/2}U_Ne^{-\beta_R H_N/2}\ket{j}.
\end{align} 
As is usual, we can understand some typical properties of the pure state $\ket{\psi_N}$ by averaging
over the choice of $U_N$.\footnote{This averaging is done with the standard formulas
$\langle U^i_j {U^\dagger}^k_l\rangle=\frac{1}{L} \delta^i_l\delta^k_j$ and $\langle U^i_j{U^\dagger}^k_l U^m_n {U^\dagger}^p_q\rangle=\frac{1}{L^2}\left(\delta^i_l\delta^k_j\delta^m_q\delta^p_n+
\delta^i_q\delta^p_j \delta^m_l\delta ^k_n \right)+ {\mathcal O}\left(\frac{1}{L^3}\right) $ for the averages over the Haar measure of an $L\times L$ unitary
matrix $U$ \cite{weingarten1978asymptotic, collins2003moments}. (In our application, $L=2^N$.) Averages of a product with an unequal number of $U$'s and $U^\dagger$'s vanish.}

There are correlations between the matrix element and the normalization factor in the formula (\ref{normalizedstate})
for $M_{ij}$ because they depend on the same unitary $U_N$. Nevertheless, it is clear that $M_{ij}$ is zero on average because $U_N$ is drawn from an ensemble that is invariant under $U_N \rightarrow -U_N$, which flips the sign of the numerator but not the denominator, so $\overline{M_{ij}} = -\overline{M_{ij}}=0$.

Since this argument does not directly extend to the gravitational case studied in the next section, we will
explain a replica trick to get the same result. First we define
\begin{align}
 M_{ij;n} = \mathcal{N}^{n}\bra{i}e^{-\beta_L H_N/2}U_Ne^{-\beta_R H_N/2}\ket{j}.
\end{align} 
The idea is to compute the average of $M_{ij;n}$ for positive integer $n$ and then analytically continue to $n=-1/2$
to get the average of the original $M_{ij}$. However, the average of $M_{ij;n}$ vanishes for any positive
integer $n$, because, as $\mathcal N$ is bilinear in $U_N$ and $U_N^\dagger$, $M_{ij;n}$ is a polynomial in
$U_N$ and $U_N^\dagger $ with 
 an unequal number of $U_N$'s and $U_N^{\dagger}$'s. Therefore, $\overline{M_{ij}} = 0$. 
 
We also note that 
\begin{align}
 \bra{\psi_N}\psi_N\rangle = \sum_{ij}|M_{ij}|^2 = 1,
\end{align}
the ensemble average of which is, of course, $1$. Since the state $\Psi_N$ is normalized but has vanishing average,
its coefficients $M_{ij}$ have large fluctuations. The fluctuations must affect the phase of $M_{ij}$, since
the average of $M_{ij}$ vanishes, but from what we have said so far, it is not clear if $M_{ij}$ fluctuates
in magnitude.

To address this, we may consider the expectation value of a simple operator $a_{ij} \equiv \ket{i}\bra{i}_L \otimes \ket{j}\bra{j}_R $ :
\begin{align}\label{ratio}
 \omega_N(a_{ij}) \equiv \bra{\psi_N}a_{ij} \ket{\psi_N} = |M_{ij}|^2 =\frac{| \bra{i} e^{-\beta_L H_N/2}U_Ne^{-\beta_R H_N/2}\ket{j}|^2}{\mathcal{N}}.
\end{align} Here there is no sum over $i$ and $j$.
Again, to average this quantity, we must consider the correlations between the numerator and denominator. In order for $\omega_N(a_{ij})$ to converge, 
the numerator and denominator on the right hand side of (\ref{ratio}) must be, in the large $N$
limit, proportional to each other. A useful criterion is that for this to be the case,
the numerator and denominator must have the same normalized variance. The variance of a random
variable $X$ is as usual $\var(X)=\langle X^2\rangle-\langle X\rangle^2$, and the normalized variance
is $\var(X)/\langle X\rangle^2$.

Averaging the unitary group over the Haar measure, the normalized variance of the denominator in
(\ref{ratio}) is seen to be 
\begin{align}
 \frac{\textrm{var}(\mathcal{N}_N)}{\overline{\mathcal{N}_N}^2} = e^{-S_2(\beta_R)-S_2(\beta_L)}+ O(2^{-N})
\end{align}
where $S_2$ is the R\'enyi 2-entropy 
of the thermal density matrices $\rho_L=\tfrac{1}{Z}{\Tr e^{-\beta_L H_N}}$,
$\rho_R=\tfrac{1}{Z}{\Tr e^{-\beta_RH_N}}$. (The R\'enyi 2-entropy of a density matrix $\rho$
is defined as $S_2(\rho)=-\log \Tr\,\rho^2$, so $e^{-S_2(\rho)}= \Tr\,\rho^2$.) 
These R\'enyi entropies are finite in the large $N$ limit if the limiting algebras are of type I.\footnote{A normal state of a type I algebra can be represented by a density matrix $\rho$, which is a positive semi-definite operator in (or affiliated to)
the algebra with $\Tr\,\rho = 1$. Such an operator satisfies $0<\Tr\,\rho^2\leq 1$, implying that
$0\leq S_2(\rho)<\infty$. By contrast, von Neumann entropy can be infinite for a state of a type I algebra, because finiteness of $\Tr\,\rho$ does not imply finiteness of $\Tr\,\rho\log\rho$. $S_2(\rho)$ vanishes precisely if $\rho$ has rank 1; thus for a thermal density matrix, it vanishes precisely if
$\beta=\infty$, assuming no ground state degeneracy. } Similarly, the fluctuations in the numerator may be evaluated by averaging over the unitary group, leading to
\begin{align}
 \frac{\textrm{var}(|\bra{i} e^{-\beta_L H_N/2}U_Ne^{-\beta_R H_N/2}\ket{j}|^2)}{\overline{| \bra{i} e^{-\beta_L H_N/2}U_Ne^{-\beta_R H_N/2}\ket{j}|^2}^2}= 1-O(2^{-N}).
\end{align}
These two normalized variances are different (when $\beta_R,\beta_L <\infty$), which implies that $\omega_N(a_{ij})$ has finite variance in the large $N$ limit. Thus, the $M_{ij}$'s have finite variance in magnitude as well as phase.

We conclude that the sequence $\{\omega_N\}$ does not converge unless $\beta_L$ and $\beta_R$ are taken to infinity, corresponding to a projection onto the ground state. Nevertheless, the sequence of partial sums can still converge to the large $N$ limit of the ensemble average, which is the thermal expectation value
\begin{align}
 \lim_{N\rightarrow \infty}\frac{1}{N}\sum_{n=1}^{N}\omega_n(a_{i,N}) = \braket{a_i}_{\beta}. \label{stronglaw}
\end{align}
The $\omega_n(a_{i,N})$'s are functions of independent random variables $U_N$, but the $U_N$ are
drawn from different distributions (namely ensembles of random unitary matrices with
an $N$-dependent size). Even though the ensembles are different
what is sometimes called Kolmogorov's strong law of large numbers implies that if the limit on the left hand
side of equation (\ref{stronglaw}) exists, then this limit is unchanged if each $\omega_n(a_{i,N})$
is replaced by its average $\overline{\omega_n(a_{i,N})}=\la a_{i,N}\ra_\beta$, leading to the claim in
eqn.~(\ref{stronglaw}).

Even if the two-sided state does not converge, it is a logical possibility that the one-sided state might converge. That is, expectation values of operators of the form $a_{i} = \ket{i}\bra{i}_L\otimes {1}_R$, namely
\begin{align}
 \omega_N(a_i) = \sum_{j}|M_{ij}|^2 = \frac{ \bra{i}e^{-\beta_L H_N/2 }U_N e^{-\beta_R H_N }U_N^{\dagger}e^{-\beta_L H_N/2 }\ket{i}}{\mathcal{N}},
\end{align} could converge.
The normalized variance of the numerator is
\begin{align}
 \frac{\textrm{var}(\bra{i}e^{-\beta_L H_N/2 }U_N e^{-\beta_R H_N }U_N^{\dagger}e^{-\beta_L H_N/2 }\ket{i})}{\overline{\bra{i}e^{-\beta_L H_N/2 }U_N e^{-\beta_R H_N }U_N^{\dagger}e^{-\beta_L H_N/2 }\ket{i}}^2} =e^{-S_2({\beta_R})} + O(2^{-N}),
\end{align}
which is distinct from that of the denominator whenever $\beta_L < \infty$, so the one-sided state does not converge unless it is the ground state.

It is also interesting to consider a parameter regime (small $\beta_R,\beta_L$ for a suitable choice of $H_N$) such that the R\'enyi entropy diverges as $N \rightarrow \infty$. This means that there is divergent entanglement
between the left and right subsystems in the large $N$ limit. This gives models of long wormholes. In this case, the sequence $\{\omega_N\}$ would converge in the large $N$ limit to a tensor product of thermal states for the left and right algebras.\footnote{One may also consider a random unitary ensemble that retains some correlation between the two sides, such as finite time evolution using a GUE matrix as a Hamiltonian. This is a better model of a long, but not infinitely long, wormhole. See also \cite{Magan:2025hce}.} However, the operator $a_{ij}$ and its one-sided version $a_i$ do not survive this limit; they converge to the zero operator. Operators that do converge include those acting nontrivially on a number of qubits that remains fixed as $N\to\infty$. Generically, such operators generate a type III$_1$ algebra. However, if $\beta_L = \beta_R = 0$, the resulting large $N$ algebra is type II.

\subsection{A Qubit Model of One-Sided Black Holes and Eigenstate Thermalization \label{onesidedmodel}}

We will now consider a one-sided version of this construction.
Given a Hamiltonian $H_N$ on $N$ qubits, a generic pure state that approximates a Gibbs state at inverse temperature $\beta$ is 
\begin{align}
 \ket{\Psi_N} = \mathcal{N}^{-1/2} e^{-\beta H_N/2}U_N\ket{0}, \quad \mathcal{N}_N\equiv \bra{0} U_N^{\dagger}e^{-\beta H_N}U_N\ket{0}.
\end{align}
where $\ket{0}$ is an arbitrary reference state. This is a reasonable model for a typical pure state black hole and has been previously considered in the quantum statistical mechanics literature \cite{Sugiura:2013pla}. A simple operator that survives the thermodynamic limit is one that acts non-trivially on a finite number of qubits. Its expectation value is
\begin{align}\label{omegaN}
 \omega_N(a) \equiv \bra{\psi_N} a\ket{\psi_N} = \frac{\bra{0}U_N^{\dagger}e^{-\beta H_N/2}ae^{-\beta H_N/2}U_N\ket{0}}{\mathcal{N}_N}.
\end{align}
The fluctuations of the numerator and denominator are
\begin{align}
\frac{\textrm{var}(\bra{0}U_N^{\dagger}e^{-\beta H_N/2}ae^{-\beta H_N/2}U_N\ket{0})}{\overline{\bra{0}U_N^{\dagger}e^{-\beta H_N/2}ae^{-\beta H_N/2}U_N\ket{0}}^2} &= e^{-S_2(\beta)}\frac{\braket{a_La_R}_{\TFD,2\beta}}{\braket{a}^2_{\beta}}
\\
 \frac{\textrm{var}(\mathcal{N}_N)}{\overline{\mathcal{N}_N}^2} &= e^{-S_2(\beta)}
\end{align}
Here, $\braket{a}_{\beta}$ is the thermal expectation value of $a$ while $\braket{a_La_R}_{\TFD,2\beta}$ is the expectation value in the thermofield double state\footnote{As usual, the thermofield
double state at inverse temperature $\beta$ is defined as $\Psi_{\TFD,\beta}=\frac{1}{\sqrt{\Tr\,e^{-\beta H_N}}}e^{-\beta H_N/2}$, viewed as a pure state of a doubled system.} at inverse temperature $2\beta$ with one insertion of $a$ on the left side and one on the right. The variances are equal if and only 
if they both vanish, 
which happens precisely if the R\'enyi entropy $S_2(\beta)$ diverges. In that case, $\omega_N(a)$ can be computed for
large $N$
by averaging the numerator and denominator in (\ref{omegaN}) over $U_N$; one learns that the
sequence $\{\omega_N\}$ converges to the Gibbs state at inverse temperature $\beta$. We see that typical sequences of pure states with $S_2(\beta)\to\infty$ converge in the large $N$ limit to mixed states. The divergence of the R\'enyi entropy implies that the large $N$ algebra is not type I. If we perform the GNS construction using the limiting state, we arrive at the thermofield double Hilbert space with the commutant of the algebra $\A$ acting on the ``other side.''

The same result holds for sequences of energy eigenstates of a chaotic system if we assume the eigenstate thermalization hypothesis \cite{Srednicki:1995pt}. This hypothesis asserts that the expectation value of simple operators in high energy eigenstates are equal to the microcanonical answer, $f_a(E)$, up to fluctuations that are exponentially suppressed in the microcanonical entropy, $S(E)$
\begin{align}
 \bra{E}a\ket{E} = f_a(E) + O(e^{-S(E)/2}).
\end{align}
The microcanonical entropy will diverge in the thermodynamic limit, implying that eigenstates converge to the (mixed) microcanonical state.

\section{The Large-N Limit of CFT and the Gravitational Path Integral}\label{three}

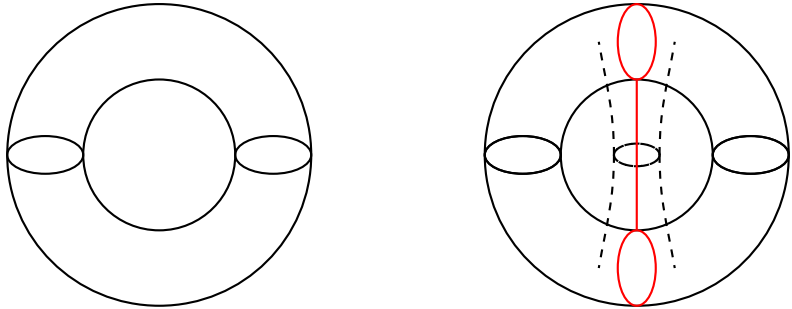
\begin{figure}
 \centering
 \begin{tikzpicture}
 \draw[thick] (0,0) arc[start angle=0, end angle=360, radius=2];
 \draw[thick] (-1,0) arc[start angle=0, end angle=360, radius=1];
 \draw[thick] (-.5,0) ellipse (.5 and .25);
 \draw[thick] (-3.5,0) ellipse (.5 and .25);
 \end{tikzpicture}
 \hspace{2cm}
 \begin{tikzpicture}
 \BUket{0}{0}{0}{1}{0}{0}
 \BUket{3}{0}{0}{-1}{0}{0}
 \draw[thick,red] (1.5,-1.5) ellipse (.25 and .5);
 \draw[thick,red] (1.5,1.5) ellipse (.25 and .5);
 \draw[thick,red] (1.5,-1) -- (1.5,1);
 \end{tikzpicture}
 \caption{Left: the path integral that computes the norm of the state when no heavy operator is inserted. Cutting on the horizontal ellipses and continuing to Lorentzian signature leads to thermal AdS if the temperature is sufficiently low. Right: the path integral with $\OO_N$ and its adjoint inserted (red ellipses). The gravitational saddle is proposed to include a baby universe (dashed line) supported by a matter shell (red line).}
 \label{fig:tAdS_norm}
\end{figure}

\subsection{Variance of the CFT State in the AS$^2$ Construction}

In the AS$^2$ construction of baby universes entangled with asymptotically AdS spacetime, 
one considers a family of CFT's labeled by an integer $N$. These
CFT's are assumed to
have a large $N$ limit that has a bulk dual description of the familiar kind. In this family of CFT's, 
one considers a family of operators
$\OO_N$ and a corresponding family of (unnormalized) states
\be\ket{\hat\Psi_N}= e^{-\beta_L H_N/2} \OO_N e^{-\beta_R H_N/2},\ee
where $H_N$ is the CFT Hamiltonian.
The normalized state is of course $\Psi_N=\hat\Psi_N/\sqrt{\la\hat\Psi_N|\hat\Psi_N\ra}$, where
\be\label{normstate}\langle\hat\Psi_N\ket{\hat\Psi_N}=\Tr\left( e^{-\beta_L H_N} \OO_N e^{-\beta_R H_N} \OO^\dagger_N\right).\ee As in section \ref{babymodel}, we view $\hat\Psi_N$ and $\Psi_N$ as pure states of
a two-sided system.
The operators $\OO_N$ will play the 
role of the random unitary $U_N$ of section \ref{two}, and we will ask about the family of
states $\Psi_N$ the same questions that we asked of certain qubit models in section \ref{two}.

The inverse temperatures $\beta_L$ and $\beta_R$
are assumed to be large enough so that if $\OO_N=1$, the state $\hat\Psi_N$ is below the Hawking-Page
transition \cite{Hawking:1982dh}, so that in the dual gravitational description, the norm of the state $\hat\Psi_N$ is
dominated by a thermal AdS space (left of fig.~\ref{fig:tAdS_norm}). The operators $\OO_N$
are assumed to depend on $N$ in some sort of regular fashion, and the idea in the AS$^2$ construction
is to pick the $\OO_N$ in such a fashion that the dual path integral computing the norm of
the state $\hat\Psi_N$ is dominated by a spacetime with a ``wormhole'' that connects the $\OO_N$
and $\OO^\dagger_N$ insertions in the formula (\ref{normstate}), as in the right of figure 
\ref{fig:tAdS_norm}. If that is the case, then the state $\hat\Psi_N$
has a gravitational dual description, obtained by making a horizontal ``cut'' in the right of fig.~\ref{fig:tAdS_norm},
as an entangled state of two asymptotically AdS regions and a baby universe.

We should note at the outset that it is far from clear what sort of operator sequences $\OO_N$ do
create a baby universe in this sense, and even whether such sequences exist.\footnote{Immediate challenges are competing saddles where the particles created by $\OO$ self annihilate in the bra and ket individually without need for a wormhole. If one tries to avoid this by giving $\OO$ some large charge, 
one runs into the problem that conserved charges in gravity are gauge charges, and particles carrying a gauge charge cannot go through the wormhole, as the total charge in a closed universe always vanishes.
It would be very interesting to obtain a clear construction where the baby universe saddle dominates.} However, in this
article, we will simply assume that we are given a sequence $\OO_N$ that leads to the wormhole
picture of fig.~\ref{fig:tAdS_norm}, and ask the following question: In such a case, does the
state $\Psi_N$ have a large $N$ limit in the boundary CFT description? We will argue that
whenever the gravitational description is dominated by a wormhole spacetime, as in the figure,
the CFT state does not have a large $N$ limit. Indeed, this is a generic feature of gravitational saddles with baby universes.

We can expand $\hat\Psi_N$ in a complete basis $|i_N\rangle$ of CFT states:
\be\label{expstate}\ket{\hat\Psi_N}=\sum_{i_N,j_N} \ket{i_N} \bra{i_N}e^{-\beta_L H_N/2} \OO_N e^{-\beta_R H_N/2}
\ket{j_N}\bra{j_N}. \ee
For the state $\hat\Psi_N$ to have a large $N$ limit, the states $\ket{i_N}$, $\ket{j_N}$ that appear in 
this expansion must have large $N$ limits. The obvious states that have large $N$ limits
are the states created by acting on the CFT ground state with finitely many single-trace operators.
For the moment we will consider states of this type; in section \ref{generalsemi}, we will 
describe a certain generalization and explain why it is relevant, but ultimately does not change the main
conclusions. 

\begin{figure}
 \centering
 \begin{tikzpicture}
 \draw[thick] (0,0) arc[start angle=0, end angle=-180, radius=2];
 \draw[thick] (-1,0) arc[start angle=0, end angle=-180, radius=1];
 \draw[dotted,thick] (-.5,0) ellipse (.5 and .25);
 \draw[dotted,thick] (-3.5,0) ellipse (.5 and .25);
 \draw[thick] (-3,0) arc[start angle=0, end angle=180, radius=.5];
 \draw[thick] (0,0) arc[start angle=0, end angle=180, radius=.5];
 \draw[thick,red] (-2,-1.5) ellipse (.25 and .5);
 \filldraw[blue] (-.5,.5) circle (2pt);
 \filldraw[blue] (-3.5,.5) circle (2pt);
 \node[blue] at (-3.5,.75) {$i$};
 \node[blue] at (-.5,.75) {$j$};
 \end{tikzpicture}
 \caption{The CFT path integral for $\hat M_{ij}$. The blue dots represent insertions of operators in the single-trace algebra.}
 \label{fig:Mij}
\end{figure}
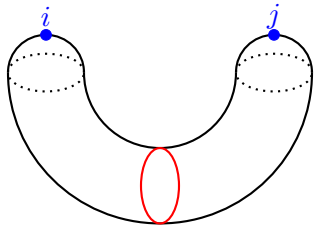

Once we assume that the states $\ket{i_N}$, $\ket{j_N}$ have large $N$ limits, it follows
that the state $\hat\Psi_N$ has a large $N$ limit if and only if the coefficients
\be\label{coeff} \hat M_{i_Nj_N} =\bra{i_N}e^{-\beta_L H_N/2} \OO_N e^{-\beta_R H_N/2}
\ket{j_N} \ee
have large $N$ limits. The coefficients $\hat M_{i_Nj_N}$ are computed,
according to AdS/CFT duality, by the path integral depicted in fig.~\ref{fig:Mij}. The
external states $\bra{i_N}$ and $\ket{j_N}$ can each be represented in the boundary
CFT by a hemisphere with a suitable operator inserted at its center. With our assumption
about the relevant states, the operators that must be inserted are elements of the algebra
of single-trace operators. A gravitational computation of the coefficient $\hat M_{i_N j_N}$
proceeds, as usual, by summing over all ways of filling in the boundary geometry of figure
\ref{fig:Mij} with a bulk geometry. 

The norm of the state $\hat\Psi_N$, on the other hand, can be expressed as a sum of absolute values
squared of the coefficients $\hat M_{i_N j_N}$:
\be\label{normstate2} \bra{\hat \Psi_N}\hat\Psi_N\rangle=\sum_{i_N,j_N} |\hat M_{i_Nj_N}|^2. \ee
Because the right hand side is bilinear in $\hat M_{i_N j_N}$ and its complex conjugate,
it is in the language of wormhole theory a ``two copy'' quantity, computed in AdS/CFT duality
by specifying two copies of the boundary (or more precisely a copy and its mirror image with hermitian 
adjoint operator insertions,
accounting for the complex conjugation of one factor of $\hat M_{i_N j_N}$). However, the hypothesis
that the gravitational path integral that computes the norm is dominated by a wormhole geometry 
means precisely that the dominant contribution to 
$\sum_{i_N,j_N}\overline{|\hat M_{i_N j_N}|^2}$ is connected. 
The path integral that computes $\overline{\hat M_{i_N j_N}}$ or its complex conjugate individually 
does not receive a contribution from the wormhole. Here, the overline is used to denote the answer from the gravitational path integral, which is interpreted as some sort of ensemble average.

Under standard assumptions about the interpretations of wormholes in gravity,
this means that the $\hat M_{i_N j_N}$, or at least some of them, behave as random variables, with a 
nontrivial variance.
Actually, if for some choice of operators $\OO_N$, the AS$^2$ construction works as claimed,
then each of the $\hat M_{i_N j_N}$ behaves as a random variable. To compute the
absolute value squared $|\hat M_{i_N j_N}|^2$ of a particular coefficient in the expansion of $\hat\Psi_N$, as opposed to the sum in (\ref{normstate2}),
 we should insert in the path integral projection operators
$\Pi^L_{i_N}$, $\Pi^R_{j_N}$ that project the left or right asymptotically AdS region onto the states $i_N$, $j_N$
and
compute not the norm $\la\hat\Psi_N |\hat\Psi_N\ra$ but the matrix element
\be\label{matrixemt} \braket{\hat\Psi_N|\Pi^L_{i_N}\Pi^R_{j_N}|\hat\Psi_N}. \ee
According to standard analyses of the AS$^2$ construction, if it is true that a wormhole geometry
dominates the computation of the norm, then, at least for sufficiently
large $\beta_L$ and $\beta_R$, insertion of the projection operators
does not affect this statement. The reason is that for large $\beta_L$ and $\beta_R$, the 
projection operators are inserted far from the operators $\OO_N$, $\OO_N^\dagger$, in regions
in which the geometry is very nearly that of empty AdS spacetime, as indicated in fig.~\ref{fig:Mij}.
Inserting the projection operators does not affect the wormhole geometry in the classical limit, though it definitely does affect the quantum corrections.

If it is true that the wormhole geometry has a large $N$ limit,
as claimed in analyses of
the AS$^2$ construction, then the variance of the coefficients $\hat M_{i_N,j_N}$ survives
in the large $N$ limit, and the limiting variables, which we will call simply $\hat M_{i j}$, are
again random variables with variance computed from the wormhole geometry. 
So the family of two-sided CFT 
states $\ket{\Psi_N}$ does not have a large $N$ limit.

We have made this analysis for the unnormalized state $\hat\Psi_N$, and we should ask whether normalizing the state
would change the answer.
The normalized matrix elements $M_{i_Nj_N}=\hat M_{i_N,j_N}/\sqrt{\hat\Psi_N|\hat\Psi_N\ra}$ 
 can be studied by a replica trick as in section \ref{babymodel}. We write 
\begin{align}
M_{i_N j_N}= \frac{\hat M_{i_Nj_N}}{\bra{\hat\Psi_N}\hat\Psi_N\rangle^{1/2}} = \lim_{n\rightarrow -1/2}\bra{\hat\Psi_N}\hat\Psi_N\rangle^{n}\hat M_{i_Nj_N}.
\end{align}
We see that for every positive integer value of $n$, the quantity on the right hand side
has $n+1$ $\OO_N$'s and $n$ $\OO_N^\dagger$'s. So 
assuming that the two-boundary wormhole of AS$^2$ is the relevant one, the quantity
$\la\hat\Psi_N|\hat\Psi_N\ra^n \hat M_{i_Nj_N}$ will receive its dominant contribution
from a contribution with $n$ disjoint copies of the AS$^2$ wormhole (each accounting for one $\OO_N$ and and one 
$\OO_N^\dagger$) and one disjoint
single-boundary contribution (accounting for the last $\OO_N$). Continuing this
statement to $n=-1/2$, we learn that the expectation value $\overline{\hat M_{i_N j_N}/\la\hat \Psi_N,\hat\Psi_N\ra^{1/2}}$ 
is proportional to a one-boundary contribution to $\hat M_{ij}$ divided by the square root of an AS$^2$ wormhole
contribution. By the assumption that the wormhole dominates the computation of $\la\hat\Psi_N |\hat\Psi_N\ra$,
this ratio is exponentially small for large $N$ and 
thus the average of the normalized state $\Psi_N$ is likewise extremely small. One could
try to avoid this conclusion by assuming an important role of (for example) a three-boundary wormhole with 
fortuitous properties, but this would be rather contrived.

\begin{figure}
 \centering
 \begin{tikzpicture}
 \begin{scope}[shift={(0,0)}, yscale=1]
 \draw[thick] (0,0) ellipse (.5 and .25);
 \draw[very thick,green] (1.5,-.25) ellipse (.3 and .15);
 \draw[thick] (3,0) ellipse (.5 and .25);
 \draw[thick] (3.5,0) arc[start angle=0, end angle=-180, radius=2];
 \draw[thick] (2.5,0) arc[start angle=0, end angle=-180, radius=1];
 \draw[thick, dashed] (1.2, -.25) to[out = -90, in =80] (1,-1.5);
 \draw[thick, dashed] (1.8, -.25) to[out = -90, in =100] (2,-1.5);
 \end{scope}%
 \begin{scope}[shift={(0,0)}, yscale=-1]
 \draw[thick] (0,0) ellipse (.5 and .25);
 \draw[thick,double] (1.5,-.25) ellipse (.3 and .15);
 \draw[thick] (3,0) ellipse (.5 and .25);
 \draw[thick] (3.5,0) arc[start angle=0, end angle=-180, radius=2];
 \draw[thick] (2.5,0) arc[start angle=0, end angle=-180, radius=1];
 \draw[thick, dashed] (1.2, -.25) to[out = -90, in =80] (1,-1.5);
 \draw[thick, dashed] (1.8, -.25) to[out = -90, in =100] (2,-1.5);
 \end{scope}%
 \begin{scope}[shift={(5,0)}, yscale=1]
 \draw[thick] (0,0) ellipse (.5 and .25);
 \draw[thick, double] (1.5,-.25) ellipse (.3 and .15);
 \draw[thick] (3,0) ellipse (.5 and .25);
 \draw[thick] (3.5,0) arc[start angle=0, end angle=-180, radius=2];
 \draw[thick] (2.5,0) arc[start angle=0, end angle=-180, radius=1];
 \draw[thick, dashed] (1.2, -.25) to[out = -90, in =80] (1,-1.5);
 \draw[thick, dashed] (1.8, -.25) to[out = -90, in =100] (2,-1.5);
 \end{scope}%
 \begin{scope}[shift={(5,0)}, yscale=-1]
 \draw[thick] (0,0) ellipse (.5 and .25);
 \draw[very thick, green] (1.5,-.25) ellipse (.3 and .15);
 \draw[thick] (3,0) ellipse (.5 and .25);
 \draw[thick] (3.5,0) arc[start angle=0, end angle=-180, radius=2];
 \draw[thick] (2.5,0) arc[start angle=0, end angle=-180, radius=1];
 \draw[thick, dashed] (1.2, -.25) to[out = -90, in =80] (1,-1.5);
 \draw[thick, dashed] (1.8, -.25) to[out = -90, in =100] (2,-1.5);
 \end{scope}%
 \end{tikzpicture}
 \caption{The path integral that gives the variance of the norm of $\hat\Psi_N$. The green surfaces are identified and the double line surfaces are identified. This is the same path integral that computes the second R\'enyi entropy.}
 \label{fig:normvariance}
\end{figure}
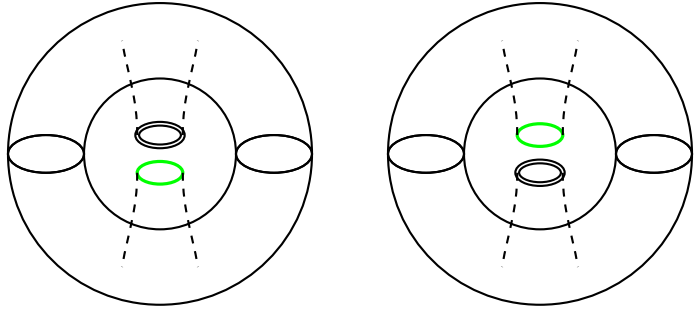

\subsection{A Four-Boundary Quantity}

This analysis makes it clear that there is variance in the phase of the normalized matrix element
$M_{ij}= \hat M_{ij}\la\hat\Psi_N|\hat\Psi_N\ra^{-1/2}$, since the average value of this quantity is small,
but does not make clear if there is variance in the magnitude of the matrix element.
As in the qubit model, by considering a four copy quantity, we can demonstrate variance in the magnitude of 
the normalized matrix elements. We proceed precisely as in section \ref{babymodel}. We will
show that 
\be\label{longone} |M_{ij}|^2 = \frac{|\hat M_{ij}|^2}{\la\hat\Psi_N|\hat\Psi_N\ra} \ee
has large variance by showing that the numerator and denominator in the expression on the right
hand side of eqn.~(\ref{longone}) have different variance.

First, we evaluate the fluctuations in the norm of $\la\hat\Psi_N |\hat\Psi_N\ra$. The expectation value of
$\la\hat\Psi_N |\hat\Psi_N\ra^2$ 
is computed in bulk by a gravitational path integral with asymptotic boundary consisting of two copies of the
boundary geometry used in computing $\la\hat\Psi_N|\hat\Psi_N\ra$. To make a connected geometry with that boundary behavior
that will contribute
to the variance of $\la\hat\Psi_N |\hat\Psi_N\ra$, we start with two copies of the AS$^2$ geometry,
 cut the baby universe in each copy on its midplane, and the glue the above-the-cut boundary of one to the below-the-cut boundary of the
other and vice-versa (fig.~\ref{fig:normvariance}). The resulting geometry has another interpretation. The left half of fig.~\ref{fig:normvariance} can be viewed as a path integral that computes an unnormalized density matrix $\widetilde\rho_{BU}$ of the baby universe. The normalized density matrix of the baby universe is
\be\label{welnorm} \rho_{BU}=\frac{\widetilde\rho_{BU}}{\la\hat\Psi_N |\hat\Psi_N\ra}. \ee
Gluing together two copies of the geometry that computes $\widetilde\rho_{BU}$, as in fig.~\ref{fig:normvariance},
gives a path integral that computes
\be\label{zelmore}\Tr\,\widetilde \rho_{BU}^2={\Tr\,\rho_{BU}^2}{\la\hat\Psi_N \hat\Psi_N\ra^2}. \ee
The normalized variance of $\la\hat\Psi_N |\hat\Psi_N\ra$ is thus $\Tr\,\rho_{BU}^2$.
This is equivalently $e^{-S_2}$, where $S_2$ is the R\'enyi entropy
of the baby universe or equivalently (as the overall state is pure) of the union of the two asymptotically AdS regions.
In particular, $S_2$ is generally assumed in the AS$^2$ construction to be finite and nonzero even in the large
$N$ limit, so the normalized variance of $\la\hat\Psi_N|\hat\Psi_N\ra$ is nonzero, but less than 1, in that limit. 

\begin{figure}
 \centering
 \begin{tikzpicture}
 \BUket{0}{0}{0}{.5}{1}{1}
 \BUket{0}{1}{1}{.5}{1}{1}
 \BUket{3}{0}{0}{.5}{2}{1}
 \BUket{3}{1}{1}{.5}{2}{1}
 \BUket{8}{0}{0}{.5}{1}{1}
 \BUket{8}{1}{1}{.5}{2}{1}
 \BUket{11}{0}{0}{.5}{2}{1}
 \BUket{11}{1}{1}{.5}{1}{1}
 \node[] at (6.25,.5) {$+$};
 \node[] at (-.5,0) {$i$};
 \node[] at (2.5,0) {$i$};
 \node[] at (10.5,0) {$i$};
 \node[] at (7.5,0) {$i$};
 \node[] at (-.5,1) {$i$};
 \node[] at (2.5,1) {$i$};
 \node[] at (10.5,1) {$i$};
 \node[] at (7.5,1) {$i$};
 \node[] at (2,0) {$j$};
 \node[] at (5,0) {$j$};
 \node[] at (13,0) {$j$};
 \node[] at (10,0) {$j$};
 \node[] at (2,1) {$j$};
 \node[] at (5,1) {$j$};
 \node[] at (13,1) {$j$};
 \node[] at (10,1) {$j$};
 \end{tikzpicture}
 \caption{The two equal wormhole contributions to $\overline{|M_{ij}|^4}$. Only the left diagram contributes to $\overline{|M_{ij}|^2}^2$. We have suppressed the operator insertions in the diagram for visual ease.}
 \label{fig:wormholeM4}
\end{figure}
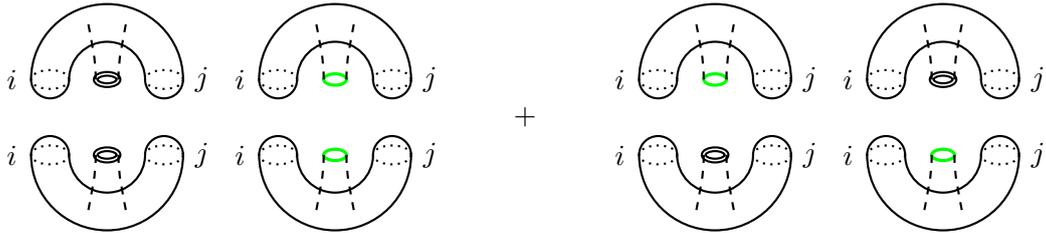

Next, we evaluate the fluctuations in $|\hat M_{ij}|^2$ using the gravitational path integral.
As previously described, the path integral for $\overline{|\hat M_{ij}|^2}$ is given by a saddle connecting, via a 
baby universe,
the
disconnected boundary manifolds appropriate to the evaluation of $\hat M_{ij}$ and of $\hat M_{ij}^{*}$. Let us call this contribution $Z_{BU}$. Its square is $Z_{BU}^2=\left(\overline{|\hat M_{ij}|^2}\right)^2$. In contrast, $\overline{|\hat M_{ij}|^4}$ involves a computation with a boundary that has four components, two appropriate for $\hat M_{ij}$ and two for 
$\hat M_{ij}^*$ (see fig.~\ref{fig:wormholeM4}).
Either of the two boundaries associated to $\hat M_{ij}$ can be connected to either of the two boundaries
associated to $\hat M_{ij}^*$ by an AS$^2$ wormhole. Summing over the two possibilities, and assuming
that there are not important connected geometries with three or four boundaries, we get
$\overline{|\hat M_{ij}|^4}=2 Z_{BU}^2=2\left(\overline{|\hat M_{ij}|^2}\right)^2$. The normalized variance of $|\hat M_{ij}|^2$ is therefore
1 in this approximation.
 Because this is generically different from the normalized variance of $\la\hat\Psi_N|\hat\Psi_N\ra$, we conclude 
 finally that there are large fluctuations in the magnitude of the coefficients $M_{ij}$ of the
 normalized state $\Psi_N$.

It is possible, of course, that the
family of pure state density matrices $\ket{\Psi_N}\bra{\Psi_N}$ can be, in some sense, averaged over $N$
to get a mixed state density matrix that has a large $N$ limit. As stated in the introduction, in this situation, the single-trace algebra will have a nontrivial commutant on the GNS Hilbert space, which one may formally associate to the baby universe operators.

We may also consider the state $\Psi_N$ at sufficiently high temperature such that the dominant bulk saddle is a long wormhole. In this case, no projection operators survive the large $N$ limit, so the previous discussion does not apply. The sequence will converge for an algebra of single-trace operators. This algebra is a type III$_1$ algebra, even when considering the union of the left and right algebras. This is in contrast to the large $N$ limit of the thermofield double state at these temperatures which is type III$_1$ for the left and right algebras individually but type I for the union. In the bulk, the fact that the resulting algebra is type III$_1$ may be seen by the entanglement of the two black hole exteriors with the large interior.

\subsection{Relation to the Toy Models}

Hopefully, the analogy of what we have found with the qubit model 
studied in section \ref{babymodel} is apparent. The operators $\OO_N$ play the role of $U_N$
in section \ref{babymodel}. The computation of the variance of $|M_{i_N j_N}|^2$ in the qubit model has a nonzero result due to the Haar integration equally weighting the two possible ways of connecting the bras and kets. In the gravitational path integral, there is similarly an equal weighting in the replica calculation where the connections come from the baby universe. In the qubit model, the variance is controlled by the thermal R\'enyi entropies while in the gravity theory, it is controlled by the R\'enyi entropies of the AdS regions which are assumed to be finite because the algebras are type I. An essentially identical analysis to the qubit model follows from the gravitational path integral when asking whether the density matrix on just one side can converge.

It is also possible to describe a one-sided analog of the AS$^2$ construction that potentially mirrors
the one-sided qubit model of section \ref{onesidedmodel}. For this, we take a CFT boundary state
$\ket{S_N}$, described by some boundary condition that has a semiclassical bulk description.\footnote{Just as most CFT's do not have semiclassical bulk duals, likewise in the case
of a CFT that does have such a bulk dual, most conformal boundary conditions do not have a semiclassical
bulk dual.} Then we act on this boundary state with an operator $\OO_N$ and define a one-sided state
\begin{equation}
\label{onepure} 
\ket{\hat{\chi}_N}=e^{-\beta H_N/2}\OO_N\ket{S_N}.
\end{equation} 
Essentially this construction was previously discussed in \cite{Balasubramanian:2025zey}. If $\beta$ is sufficiently small, then the sequence of states can converge, even without averaging, because the R\'enyi entropy will diverge in the large $N$ limit. The resulting state will be thermal and the large $N$ algebra type III$_1$. 

However, for sufficiently large $\beta$, the bulk path integral computing the norm might possibly be dominated by a wormhole
geometry, in which case the CFT state $\chi_N$ has a bulk dual consisting of a baby universe
entangled with a single asymptotically AdS region. 
This state can of course be expanded
in a complete set of CFT states $\ket{i_N}$:
\be\label{expure}\ket{\hat{\chi}_N}=\sum_{i_N} M_{i_N}\ket{i_N},~~~ M_{i_N}=\bra{i_N}\hat{\chi}_N\rangle.\ee
As in the two-sided case, if the gravitational
computation of $\overline{\bra{\hat{\chi}_N}\hat{\chi}_N\rangle}$ is dominated by a wormhole geometry, then this means that the coefficients
$M_{i_N}$ have a non-zero variance and the family of states $\chi_N$ does not have a large $N$ limit
as a pure state. Once again, it may be possible to define a mixed state density 
matrix that does have a large $N$ limit by averaging $\ket{\chi_N}\bra{\chi_N}$ over $N$. 

\subsection{General Semiclassical States}\label{generalsemi}

Although this will not affect any of the main conclusions, the preceding analysis really needs to be 
generalized in one essential way. In many discussions of the AS$^2$ construction, it is 
assumed that the operator $\OO_N$, in the large $N$ limit, creates an infinitely thin 
shell of matter.\footnote{This idea is nicely depicted in fig.~1 of \cite{Antonini:2024mci}.} Because the shell is infinitely thin, the ``wormhole'' solution 
can be built by gluing together two exact solutions, each of which is a portion of empty 
AdS spacetime. In that case, when we ``cut'' as in fig.~\ref{fig:tAdS_norm} to reveal the 
intermediate state, we are cutting in a region in which the geometry is 
precisely that of empty AdS space. Taking quantum corrections into account, that means that the states
that propagate through the cut are states obtained by perturbing empty AdS space by finitely
many quanta. The resulting states are obtained in the boundary CFT by acting on 
the vacuum with finitely many single-trace operators. These are the most obvious states that
have large $N$ limits, and the preceding discussion -- like much of the literature on the 
AS$^2$ construction -- was phrased assuming that these are the relevant states.

An infinitely thin shell is of course an idealization. More realistically, assuming that a 
suitable sequence of operators $\OO_N$ does create a shell of matter that propagates 
``through the wormhole'' as proposed in the literature, this shell will have a 
non-zero thickness. In that case,
the wormhole geometry cannot literally be built by gluing together two pieces of empty AdS space,
though (for large $\beta_L,\beta_R$) that can be a good approximation far from the shell. 
In particular, when we ``cut'' as in fig.~\ref{fig:tAdS_norm} to reveal the intermediate states,
we are cutting in a region in which the geometry is not quite that of empty AdS space. 
The cut can nonetheless be made on a plane of symmetry (the figure has an up-down reflection
symmetry that exchanges $\OO_N$ and $\OO_N^\dagger$), so it is possible to continue from the
cut to Lorentz signature. In Lorentz signature, in the classical limit,
the solution in each of the asymptotically
AdS regions is then not simply empty AdS but differs from this by the presence of some sort of coherent
classical wave. Such a wave has an energy of order $1/G$, which from the point of view
of the boundary theory means that the energy is of order $N^2$ for large $N$ in the example
of AdS$_5\times S^5$ (in some other examples, $1/G$ scales as a different power of the integer
$N$ that parametrizes the family of CFT's). To the extent that the shell is thin and
$\beta_L,\beta_R$ are large, the coefficient of $N^2$ is small, but it is nonzero and positive.
So the energy of the relevant states diverges for $N\to\infty$. In the literature, projection operators onto the low energy ($O(N^0)$) Hilbert space has been applied to remove high-energy ``tails.'' Away from the infinitely thin shell limit such operators project onto to an exponentially small branch of the wavefunction and so dramatically change the state.

Thus the relevant states are not really the most obvious states that have large $N$ limits,
namely the states -- created from the ground state by acting with finitely many 
single-trace operators -- whose energy remains finite for $N\to\infty$. 
Rather, we are running into more generic states
that have large $N$ limits. Given any classical solution of a bulk gravity 
theory in Lorentz signature -- 
including solutions that if evolved to the future or past will develop black hole or 
white hole singularities -- one can construct a Hilbert space to all orders in $G$ by quantizing small
fluctuations around this classical solution. From the point of view of the boundary CFT,
the states in this Hilbert space have large $N$ limits (and a systematic $1/N$ expansion to all orders).
Conversely, it is plausible that a construction of this type gives the most general 
possible construction of an $N$-dependent family of states that has a large $N$ limit.
In the special case of this construction in which the classical solution considered is empty
AdS space, the states obtained by this construction are the usual finite energy states obtained
by acting on the ground state with finitely many single-trace operators. In the case of a general
classical solution, the states obtained this way form an irreducible representation of the algebra
of subtracted\footnote{In the presence of a non-trivial classical solution, the single-trace
operators have expectation values that grow as a positive power of $N$; these expectation values
have to be subtracted to define an algebra of single-trace operators that makes sense in the
Hilbert space obtained by quantizating fluctuations around the given solution.} single-trace operators.

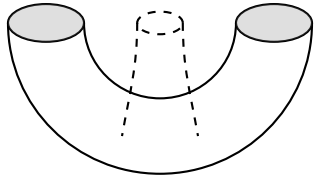
\begin{figure}
 \centering
 \begin{tikzpicture}
 \BUket{0}{0}{0}{1}{0}{0}
 \fill[gray, opacity = .25] (0,0) ellipse (.5 and .25);
 \fill[gray, opacity = .25] (3,0) ellipse (.5 and .25);
 \end{tikzpicture}
 \caption{For the general case, projection operators can be constructed by placing boundary conditions in the bulk corresponding to the relevant classical solution (shaded gray) and inserting single-trace operators at the boundary.}
 \label{fig:bulkprojector}
\end{figure}

Thus, in the preceding discussion, we really should consider the states $\ket{i_N}$ and
$\ket{j_N}$ to be states with large $N$ limits that are obtained by quantizing around some 
non-trivial classical solution. However, this does not affect any of the main points in the 
preceding derivation. The only real difference is that the projection operators $\Pi^L_{i_N}$,
$\Pi^R_{j_N}$ cannot be described by the procedure of fig.~\ref{fig:Mij}. In fact, in general
there is no convenient boundary construction of these projection operators. However, there is no
problem with a bulk construction of the projection operators, as in fig.~\ref{fig:bulkprojector}. One simply
puts a boundary condition on the bulk geometry that singles out the classical solution that 
we actually want,\footnote{The obvious way to try to do that would be to use Dirichlet boundary
conditions, fixing the metric and other fields on the boundary to be whatever one gets in cutting
the wormhole geometry that dominates the computation of the norm. But technically, in the
context of gravity, it is better to specify the conformal structure of the
boundary and the trace of the extrinsic curvature \cite{Avramidi:1997sh,Anderson:2006lqb,Anderson:2007jpe,Witten:2018lgb}.} 
embroidered with any number of single-trace 
operators acting on the boundary. Since the relevant 
projection operators select precisely the geometry that is present anyway in the classical
limit, their insertion does not affect the classical limit of the geometry and so everything that
we said previously about the large $N$ behavior of the states $\Psi_N$ remains valid,
even though the assumption that the asymptotically AdS regions are empty
in the classical limit was oversimplified.

\section{Discussion}
\label{four}

Typically, in the AdS/CFT correspondence, semiclassical gravity
 corresponds to a large $N$ limit of the boundary CFT. This assertion is difficult to interpret in
the context of the AS$^2$ construction, in which the CFT state does not have a large $N$ limit. The closest one
can come is to consider the average over $N$ of the CFT state $\Psi_N$, which can exist as a mixed state. 
Possibly what happens is that for a suitable family of operators $\OO_N$, the AS$^2$ construction does produce
a bulk geometry including a wormhole and this geometry is dual to the average over $N$ of the CFT state.
See also \cite{Liu:2025cml} for an independent discussion of averaging over $N$ as a possible interpretation 
of the AS$^2$ construction.

In general, wormholes appear to produce nonlocal effective actions in spacetime, and it has been argued, originally in
\cite{Coleman:1988cy}, that this should be interpreted in terms of averaging over the values of coupling parameters.
Averaging over quite a few parameters seems to be
required, possibly as many as the dimension of a baby universe Hilbert space, as further discussed in 
\cite{Maldacena:2004rf}. From an AdS/CFT standpoint, this is puzzling, since we seem
to have well-defined quantum gravity theories that depend on only a small number of parameters, in some cases only
a single positive integer $N$. This motivated a suggestion that perhaps averaging over $N$ is sufficient to account
for wormhole physics \cite{Schlenker:2022dyo}. Unfortunately, simple averaging over $N$ does not appear to
give satisfactory results, and a suitable alternative is not clear. 

To see the problem with simple averaging
over $N$, consider an observable $\Theta$ that depends on $N$. Suppose that, because of wormholes,
$\Theta$ behaves as a random variable
with some distribution. For simplicity assume that this distribution is independent of $N$ and that there is no
correlation between different values of $N$. We thus have a sequence of independent random values $\Theta_1,\Theta_2,\cdots$.
Ideally from this sequence we want to extract a $1/N$ expansion
\be\label{1overNexp} \Theta_N= a_0+\frac{a_1}{N}+\frac{a_2}{N^2}+\cdots \ee
which hypothetically is the answer that will come from perturbation theory in $G$. Here for definiteness we are
assuming a model and a variable $\Theta$ such that the $1/N$ expansion is expected to take the form suggested in eqn.
(\ref{1overNexp}). From the random sequence $\Theta_1,\Theta_2,\cdots$, there is no problem in extracting a 
candidate value of the leading coefficient $a_0$:
\be\label{tellme}a_0=\lim_{N\to \infty} \frac{1}{N}\sum_{m=1}^N \Theta_m . \ee
But it is difficult to see how one could possibly extract the higher order terms in the $1/N$ expansion. Indeed,
the quantity $a_0(N)=\frac{1}{N}\sum_{m=1}^N \Theta_m$ will typically differ from its large $N$ limit $a_0$
by an amount of order $1/\sqrt N$. The coefficient of $1/\sqrt N$ does not have a limit as $N$ increases,
and the existence of fluctuations of this magnitude appears to obstruct an attempt to define the subleading
coefficients $a_1,a_2,\cdots$ in the $1/N$ expansion (\ref{1overNexp}). 

Going back to our specific problem of the AS$^2$ construction, we can certainly define an average density matrix
\be\label{avstate}\rho_{\mathrm{av}}=\lim_{N\to\infty} \frac{1}{N} \ket{\Psi_N}\bra{\Psi_N}\ee
assuming that this limit converges.
But as just explained, it is not obvious how to define $1/N$ corrections to this average density matrix.

An alternative possibility, as opposed to averaging over $N$, is that
the fluctuations in the expectation values of single trace observables may be telling us
that in the AS$^2$ construction, there is no simple, natural sequence of operators $\OO_N$ (this was also
considered in \cite{Liu:2025cml}).   Rather,
there may be many sequences that are hard to distinguish, with no way to pick out a preferred sequence.
Each sequence of operators would lead to a sequence of states with the same average density matrix $\rho_{\mathrm {av}}$. In the example of one-sided black holes, this is the statement that the state at each value of $N$ is just one of $e^{S(E)}$ states in the microcanonical energy band that will lead to the same limiting (mixed) state. In the AS$^2$ setting, 
it is proposed that the solution can be described by a matter shell and, in some approximation, only depends on the 
stress-energy of the matter shell.
 It is then reasonable to expect, in analogy to the black hole setting, that the number of distinct operators leading to the same gravity solution is exponential in some power of $N$. The gravitational path integral may then be interpreted as averaging over all such choices that lead to the same average density matrix. Thus, by averaging over macroscopically indistinguishable operators, the fluctuations in the density matrix could be made to be exponentially small in $N$, which would make a $1/N$ expansion of the density matrix well-defined.

 A final comment concerns the ``meaning'' of the baby universe in the AS$^2$ construction.   
 In the large $N$ limit of this construction, the CFT state
 is a mixed state for the algebra $\A$ of single-trace operators and therefore this algebra has a commutant, which 
 conjecturally is the algebra of operators in the baby universe.  
 However, this commutant
 is an emergent notion at large $N$ and only makes sense order by order in a $1/N$ expansion. For any definite 
 integer $N$, the single-trace operators generate the full CFT operator algebra and there is no commutant.
This seems to tell us that  the baby universe can only be defined in an asymptotic sense for large $N$, a puzzling claim
as the ``baby'' universe may be big enough to contain stars, galaxies, and a whole civilization.  If it is true that
from a CFT point of view, the baby universe only can be defined in an asymptotic sense for large $N$, what can possibly be
the bulk dual of that statement?   A thought that comes to mind is that actually, the baby universes that arise
in the AS$^2$ construction, because they are static on a time zero slice, are big bang/big crunch cosmologies,
with a finite, though possibly quite long, lifetime to the past and the future.   In quantum mechanics, an unstable particle is only precisely defined in an asymptotic sense, to the extent that it is long-lived, and perhaps the same
assertion holds for a baby universe.

\acknowledgments

We thank Hong Liu, Juan Maldacena, Martin Sasieta, and Brian Swingle for discussions. JKF is supported by the Marvin L. Goldberger
Member Fund at the Institute for Advanced Study and the National Science Foundation under Grant PHY-2514611. Research of EW is partly supported by NSF Grant PHY-2514611.

\bibliographystyle{JHEP}
\bibliography{main}
\end{document}